\documentclass[aps,superscriptaddress,twocolumn,nofootinbib]{revtex4}



\usepackage{graphicx}
\usepackage{dcolumn}
\usepackage{natbib}

\begin{document}


\title{What really causes large price changes?}



\author{J. Doyne Farmer}
\affiliation{Santa Fe Institute, 1399 Hyde Park Road, Santa Fe, NM 87501}

\author{L\'aszl\'o Gillemot}
\affiliation{Santa Fe Institute, 1399 Hyde Park Road, Santa Fe, NM 87501}
\affiliation{Budapest University of Technology and Economics,
H-1111 Budapest, Budafoki \'ut 8, Hungary}

\author{Fabrizio Lillo}
\affiliation{Santa Fe Institute, 1399 Hyde Park Road, Santa Fe, NM 87501}
\affiliation{Istituto Nazionale per la Fisica della Materia, Unit\`a di Palermo, Italy}

\author{Szabolcs Mike}
\affiliation{Santa Fe Institute, 1399 Hyde Park Road, Santa Fe, NM 87501}
\affiliation{Budapest University of Technology and Economics,
H-1111 Budapest, Budafoki \'ut 8, Hungary}

\author{Anindya Sen}
\affiliation{Santa Fe Institute, 1399 Hyde Park Road, Santa Fe, NM 87501}
\affiliation{Mathematics Dept., University of Chicago}
\date{\today}

\begin{abstract}
We study the cause of large fluctuations in prices in the London Stock
Exchange.  This is done at the microscopic level of individual events,
where an event is the placement or cancellation of an order to buy or
sell.  We show that price fluctuations caused by individual market
orders are essentially independent of the volume of orders.  Instead,
large price fluctuations are driven by liquidity fluctuations,
variations in the market's ability to absorb new orders. Even for the
most liquid stocks there can be substantial gaps in the order book,
corresponding to a block of adjacent price levels containing no
quotes.  When such a gap exists next to the best price, a new order
can remove the best quote, triggering a large midpoint price change.
Thus, the distribution of large price changes merely reflects the
distribution of gaps in the limit order book.  This is a finite size
effect, caused by the granularity of order flow: In a market where
participants placed many small orders uniformly across prices, such
large price fluctuations would not happen.  We show that this
also explains price fluctuations on longer timescales.  In addition, we
present results suggesting that the risk profile varies from stock to
stock, and is not universal: lightly traded stocks tend to have more extreme
risks.
\end{abstract}


\maketitle
\tableofcontents

\section{Introduction}

It has now been known for more than forty years that price changes are
fat-tailed
\cite{Mandelbrot63,Fama65,Officer72,Akigiray89,Koedijk90,Loretan94,%
  Mantegna95,Longin96,Lux96,Ghashghaie96,Muller98,Plerou99,Mantegna99},
i.e there is a higher probability of extreme events than in a normal
distribution.  This is important for financial risk, since it means
that large price fluctuations are much more common than one might
expect.  There have been several conjectures about the origin of fat
tails in prices
\cite{Clark73,Arthur97,Mandelbrot97,Sornette98,Sornette00,Tsallis04,Cont00,
  Lux99,Giardina03,Johnson03,Marsili01,Gabaix03}.  Most of these
theories are either generic mechanisms for generating power laws, such
as multiplicative noise or maximization of alternative entropies or
agent-based models that make qualitative predictions that
are not very specific.

Two theories that deserve special mention because they make testable
hypotheses about the detailed underlying mechanism are the
subordinated random process theory of Clark \cite{Clark73} and the
recent theory of Gabaix et al. \cite{Gabaix03}.  Clark's proposal is
that because order arrival rates are highly intermittent, aggregating
in a fixed time interval leads to fat tails in price returns.
Gabaix et al.'s proposal is that high volume orders cause large price
movements.  We show that neither of these theories describe large
price changes in the London Stock Exchange\footnote{For more discussion
  of the paper of Gabaix et al. see Reference~\cite{Farmer04}.}.

Instead, we show that large price fluctuations are driven by
fluctuations in liquidity, i.e. variations in the response of prices
to changes in supply and demand.  The number of agents that
participate in the market at any given time, and thus the number of
orders to buy or sell, is rather small.  Even for a heavily traded
stock, the typical number of orders on one side of the book at any
given time is generally around $30$.  While it is in some cases a good
approximation to regard the market as a statistical system, which can
be treated using mathematical methods from statistical mechanics
\cite{Daniels03,Bouchaud02,Potters03,Smith03,Farmer03,Bouchaud04},
markets are far from the thermodynamic limit, and display strong
finite size effects.  Fluctuations in orders are important, but it is
not the size of orders that drives large price changes, but rather the
uniformity of their coverage of price levels.  Revealed supply and
demand curves at any instant in time are irregular step-like functions
with long flat regions and large jumps.  The market can be regarded as
a granular medium, in which the incremental changes in supply and
demand are the grains.  The statistical properties of prices depend
more on the fluctuations in revealed supply and demand than on their
mean behavior.

A common assumption is that large price returns $r(t)$ are
asymptotically distributed as a power law.  In more technical terms,
letting $m(t)$ be the midprice at time $t$, this means that $r(t) =
\log m(t) - \log m(t - \tau)$ satisfies $P(r > x) \sim x^{-\alpha}$.
$P(r > x)$ is the probability that $r > x$, $\tau$ is an arbitrary
time interval, and $f(x) \sim g(x)$ means that $f(x)$ and $g(x)$ scale
the same way\footnote{$f(x) \sim g(x)$
  means that $\lim_{x \to \infty} L(x)f(x)/g(x) = 1$, where $L(x)$ is
  a slowly varying function.  A slowly varying function $L(x)$ is a
  positive function that for every $t$ satisfies $\lim_{x \to \infty}
  L(tx)/L(x) = 1$.} in the limit $x \to \infty$.  $\alpha$ is called the {\it tail exponent}.  It
was initially thought that $\alpha < 2$, which is significant because
this would imply that the standard deviation of price returns does not exist,
and that under aggregation independent price returns should converge
to a Levy stable distribution \cite{Mandelbrot63,Fama65}.  However,
most subsequent studies indicate that $\alpha > 2$ is more common
\cite{Officer72,Akigiray89,Koedijk90,Loretan94,%
  Mantegna95,Longin96,Lux96,Muller98,Plerou99,Mantegna99}.
Nonetheless, it still remains controversial whether a power law is
always the best description of price returns. In this paper we do not
attempt to resolve this debate.  However, we will use power laws as a
useful way to describe the asymptotic behavior of price changes, in
particular to compare the distribution of price changes for different
stocks.  Our results suggest that tail exponents vary from stock to
stock, and are not universal.

In the remainder of this section we review the continuous double
auction, explaining what we define as an ``event'', and introducing
notation that we will use throughout the rest of the paper.  We also discuss the relationship between returns and events, and review related literature.
Section~II presents some summary statistics for the data set.
Section~III demonstrates that there is very little difference in the
price response of large and small orders, and that large price
fluctuations are driven by fluctuations in liquidity.  This is made
more explicit in Section~IV, where we show how gaps in the occupied
price levels in the orderbook lead to large price changes, and show
that the gap distribution closely matches the return distribution.  In
Section~V we compare the returns triggered by market orders, limit
orders, and cancellations, and show that they are very similar.
Section~VI demonstrates that the tail behavior of returns on the event
scale matches the tail behavior on longer timescales, and that returns
in event time and real time are similar.  Section~VII studies the
behavior of the gap distribution in more detail, demonstrating that
the (at least approximate) power law behavior we observe is not driven by
fluctuations in the number of orders in the book, but rather depends
on correlations in occupied price levels.  We conclude and offer a few
speculations about the ultimate explanation of large price
fluctuations in Section~VIII.

\subsection{Background: Continuous double auction\label{background}}

To understand our results it is essential that the reader understand
the double continuous auction, which is the standard mechanism for
price formation in most modern financial markets.  Agents can place
different types of orders, which can be grouped into two
categories: Impatient traders submit \textit{market orders}, which are
requests to buy or sell a given number of shares immediately at the
best available price. More patient traders submit \textit{limit
  orders}, or {\it quotes} which also state a limit price $\pi$,
corresponding to the worst allowable price for the transaction.  (Note
that the word ``quote'' can be used either to refer to the limit price
or to the limit order itself.) Limit orders often fail to result in an
immediate transaction, and are stored in a queue called the
\textit{limit order book}. Buy limit orders are called \textit{bids},
and sell limit orders are called \textit{offers }or \textit{asks}. At
any given time there is a best (lowest) offer to sell with price
$a(t)$, and a best (highest) bid to buy with price $b(t)$.  These are
also called the {\it inside quotes} or the {\it best prices}.  The
price gap between them is called the \textit{spread} $s(t)=a(t)-b(t)$.
Prices are not continuous, but rather change in discrete quanta called
{\it ticks}, of size $\Delta p$.  The number of shares in an order is
called either its {\it size} or its {\it volume}.

\begin{figure}[ptb]
  \begin{center}
  \includegraphics[scale=0.5]{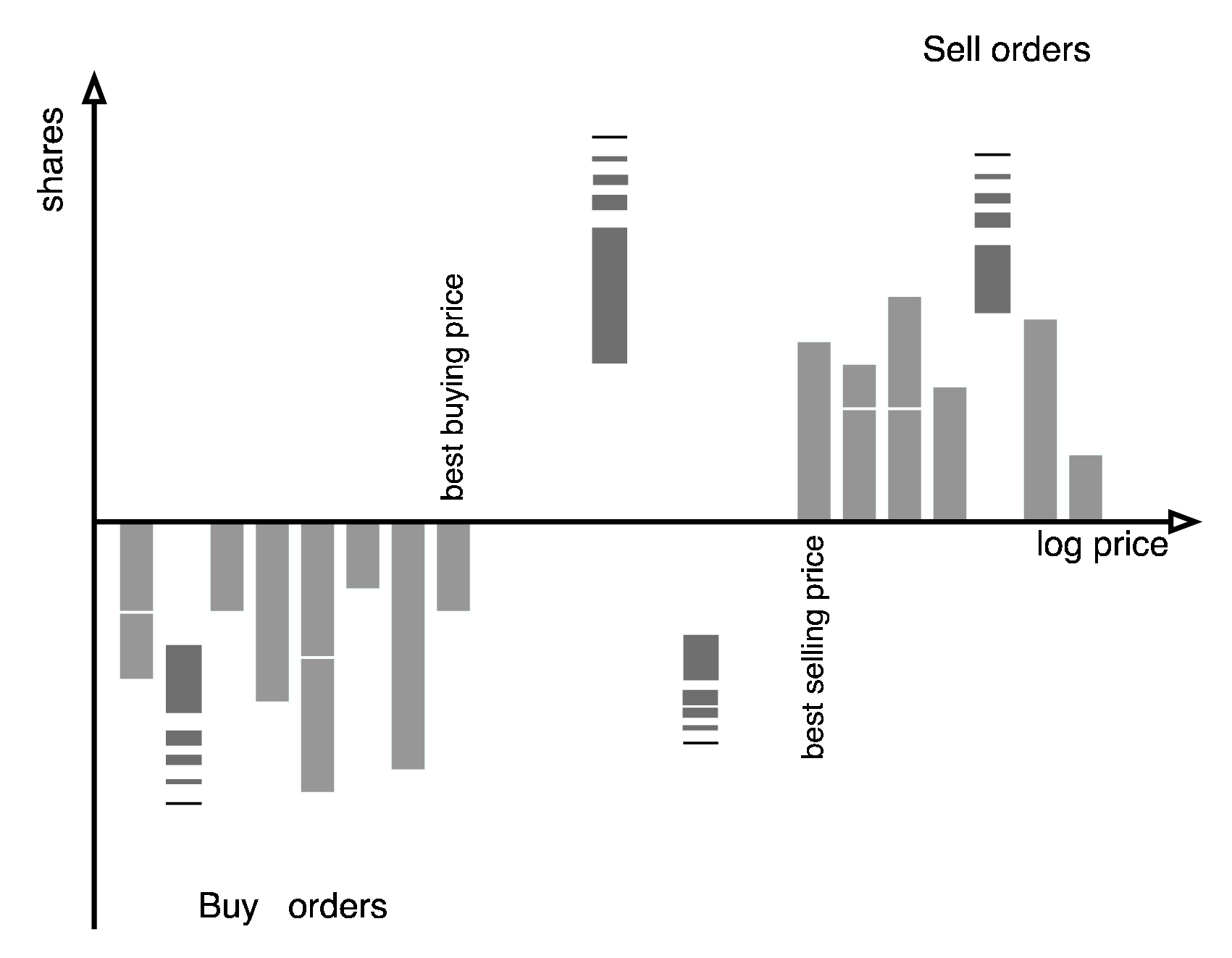} 
  \caption{ A schematic illustration of the continuous double auction
    mechanism.  Limit orders are stored in the limit order book.  We
    adopt the arbitrary convention that buy orders are negative and
    sell orders are positive.  As a market order arrives, it has
    transactions with limit orders of the opposite sign, in order of
    price (first) and time of arrival (second).  The best quotes at
    prices $a(t)$ or $b(t)$ move whenever an incoming market order has
    sufficient size to fully deplete the stored volume at $a(t)$ or
    $b(t)$, when new limit orders are placed inside the spread (when
    the limit price satisfies $b(t) <\pi < a(t)$) or when all the
    orders at the best prices are canceled.  Bids and offers that
    fall inside the spread become the new best bids and offers.}
  \label{bookschematic}
  \end{center}
\end{figure}

As market orders arrive they are matched against limit orders of the
opposite sign in order of first price and then arrival time, as shown in
Fig.~\ref{bookschematic}.
Because orders are placed for varying numbers of shares, matching is
not necessarily one-to-one. For example, suppose the best offer is for
200 shares at \$60 and the the next best is for 300 shares at \$60.25;
a buy market order for 250 shares buys 200 shares at \$60 and 50
shares at \$60.25, moving the best offer $a(t)$ from \$60 to \$60.25.
A high density of limit orders per price results in high
\textit{liquidity} for market orders, i.e., it implies a small movement
in the best price when a market order is placed.

There are a variety of different order types defined in real markets,
whose details differ from market to market.  For our purposes here,
any given order can always be decomposed into two types: We will call
any component of an order that results in immediate execution an {\it
  effective market order}, and any component that is not executed
immediately, and is stored in the limit order order book, an {\it
  effective limit order}.  For example, consider a limit order to buy
whose limit price $\pi = a(t)$.  Suppose the volume at $a(t)$ is
$1000$ shares, and the volume of the new limit order is $3000$ shares.
Then this limit order is equivalent to an effective market order for
$1000$ shares, followed by an effective limit order of $2000$ shares
with limit price $a(t)$.  In either case the same transactions take
place, and the best prices move to $b(t+1) = a(t)$ and $a(t+1) = a(t)
+ g(t)$, where $g(t)$ is the price interval to the next highest
occupied price level.  Throughout the remainder of the paper we will
simply call an effective limit order a ``limit order'', and an
effective market market order a ``market order''.

When a market order arrives it can cause changes in the best prices.
This is called {\it market impact} or {\it price impact}.  Note that
the price changes are always in the same direction: A buy market order
will either leave the best ask the same or make it bigger, and a sell
market order will either leave the best bid the same or make it
smaller.  The result is that buy market orders can increase the
midprice $m(t) = (a(t) + b(t))/2$, and sell orders can decrease
it.

There is no unique notion of price in a real market.  We will let
$\pi$ be the {\it limit price} of a limit order, and $m(t) = (a(t) +
b(t))/2$ be the {\it midpoint price} or {\it midprice} defined by the
best quotes.  All the results of this paper concern the midprice,
rather than transaction prices, but at longer timescales this makes
very little difference, since the midpoint and transaction prices
rarely differ by more than half the spread.  The midpoint price is
more convenient to study because it avoids problems associated with
the tendency of transaction prices to bounce back and forth between
the best bid and ask.  Price changes are typically characterized as
{\it returns} $r_\tau (t) = \log m(t) - \log m(t - \tau)$.

\subsection{What causes returns?}

In this paper we study changes in the midprice at the level of
individual events.  The arrival of three kinds of events can cause the
midprice to change:
\begin{itemize}
\item
{\it Market orders}.  A market order bigger than the opposite best
quote widens the spread by increasing the best ask if it is a buy
order, or decreasing the best bid if it is a sell order.
\item
{\it Limit orders}.  A limit order that falls inside the spread
narrows it by increasing the best bid if it is a buy order, or
decreasing the best ask if it is a sell order.
\item
{\it Cancellations}.  A cancellation of the last limit order at the
best price widens the spread by either increasing the best ask or
decreasing the best bid.
\end{itemize}

We prefer to study individual events for several reasons: (1) It
removes any ambiguity about time scale, and makes it easier to compare
stocks with different activity levels.  (2) It minimizes problems
associated with clustered volatility (the positive autocorrelation of
the absolute value of price changes).  Clustered volatility is also
driven by variations in event arrival rates, so its effect is weaker
when the time lag between price changes is measured in terms of number
of intervening events.  (3) Individual events are the most fundamental
level of description -- returns on any longer scale can be built out
of returns on an event scale.  We will generally measure time in terms
of {\it event time}: the time interval between events $t_1$ and $t_2$
is measured as the number of intervening events plus one.  (One is
added so that two adjacent events comprise an event time interval of
one).

Analysis at the event scale is particularly useful under the
assumption that large price fluctuations are asymptotically power law
distributed because of the invariance properties of power laws under
aggregation. If two independent variables $X$ and $Y$ both have power
law tails with exponents $\alpha_X$ and $\alpha_Y$, then the variable
$Z = X + Y$ also has a power law tail with exponent $\min (\alpha_X,
\alpha_Y)$.  Thus, if we show that returns at the level of
individual events have power law tails with exponent $\alpha$ and if
there are not strong and persistent correlations, returns on any
longer event time scale should also have power law tails with exponent
$\alpha$.  In fact, $\alpha$ appears to be the same whether returns
are measured in event time or real time (see Section~\ref{aggregation}).

\subsection{Review of previous work}

There is considerable prior work using limit order data to address
questions about market microstructure.  For example, in a very early
study Niederhoffer and Osborne discussed the granularity of revealed
supply and demand, and showed orders tend to cluster at particular
prices \cite{Niederhoffer66}.  Clustering of limit orders, as well as
stop-loss and take-profit orders is also studied in references
\cite{Kavajecz99,Osler03,Kavajecz03}.  Of particular relevance is the
work of Biais, Hillion and Spatt \cite{Biais95} who for the Paris
Bourse document several properties of order flow, including the
concentration of orders near the best prices.  They study the relation
between order flow and the dynamics of prices, and mention the
existence of gaps in the limit order book.  Another relevant
observation is made by Knez and Ready \cite{Knez96} and Petersen and
Umlauf \cite{Petersen94} who demonstrate that for the New York Stock
Exchange (NYSE) the most important conditioning variable for price
impact is the size of an order relative to the volume at the best
price.  Goldstein and Kavajecz \cite{Goldstein00} document the effect
of changes in tick size on limit order book depth, while several
theoretical papers justify the existence of positive bid-ask spread
and investigate the motivations for placing limit orders as opposed to
market orders
\cite{Cohen81,Parlour98,Foucault99,Kaniel03,Griffiths00,Hollifield02}.
In addition we should mention recent empirical studies by Coppejans
and Domowitz \cite{Coppejans02} and Coppejans, Domowitz and Madhaven
\cite{Coppejans03} that document variations in liquidity and
comovements of liquidity with returns and volatility.

Our work adds to this literature by investigating the relationship
between order placement and price movement at the level of individual
events.  Our motivation is to understand what drives large price
movements, to gain insight into the fat tails of price returns.  A
relevant paper in this regard is the work of Plerou et
al. \cite{Plerou00}, who showed that for the NYSE in a fixed time
interval the scaling behavior in the standard deviation of individual
price fluctuations roughly matches that of price fluctuations, and
dominates over fluctuations in the number of events.  This seems to
contradict the later conclusions of the same authors in Gabaix et
al. \cite{Gabaix03}.

\section{Data\label{data}}

In order to a have a representative sample of high volume stocks we
select $16$ companies traded on the London Stock Exchange (LSE) in the
4-year period 1999-2002.  The stocks we analyzed are Astrazeneca
(AZN), Baa (BAA), BHP Billiton (BLT), Boots Group (BOOT), British Sky
Broadcasting Group (BSY), Diageo (DGE), Gus (GUS), Hilton Group (HG.),
Lloyds Tsb Group (LLOY), Prudential (PRU), Pearson (PSON), Rio Tinto
(RIO), Rentokil Initial (RTO), Reuters Group (RTR), Sainsbury (SBRY),
Shell Transport \& Trading Co. (SHEL).  These stocks were selected
because they have high volume and they are all continuously traded
during the full period.  Table~\ref{summary} gives a summary of the
number of different events for the $16$ stocks\footnote{We have
  observed similar results on less liquid stocks, except that the
  statistics tend to be poorer, and as we discuss in
  Section~\ref{granularity}, e.g. Figure~\ref{scatterRtn}, less liquid
  stocks appear to have lower tail exponents.}.
\begin{table}
\caption{Summary statistics of the $16$ stocks we study for the period
  1999-2002.  The columns give the number of events of each type, in
  thousands.  All events are ``effective'' events -- see the
  discussion in Section~\ref{background}.}
\begin{tabular}{l|ccc|r}
tick & market orders & limit orders & cancellations& total\\
\tableline
AZN  &  ~652 & 2,067 & 1,454~~~ &  ~~4,173 \\
BAA  &  ~226 &  683 &  487~~~ &  ~~1,397  \\
BLT  &  ~297 &  825 &  557 ~~~&  ~~1,679  \\
BOOT &  ~246 &  711 &  501 ~~~&  ~~1,458  \\
BSY  &  ~404 & 1,120 &  726 ~~~& ~~ 2,250 \\
DGE  &  ~527 & 1,329 &  854 ~~~& ~~ 2,709 \\
GUS  &  ~244 &  734 &  518 ~~~&  ~~1,496  \\
HG.  &  ~228 &  676 &  472 ~~~&  ~~1,377  \\
LLOY &  ~723 & 1,664 & 1,020 ~~~&  ~~3,407 \\
PRU  &  ~448 & 1,227 &  821~~~ &  ~~2,496 \\
PSON &  ~373 & 1,063 &  734 ~~~& ~~ 2,170 \\
RIO  &  ~381 & 1,122 &  771 ~~~&  ~~2,274 \\
RTO  &  ~276 &  620 &  389 ~~~&  ~~1,285  \\
RTR  &  ~479 & 1,250 &  820 ~~~&  ~~2,549 \\
SBRY &  ~284 &  805 &  561 ~~~&  ~~1,650  \\
SHEL &  ~717 & 4,137 & 3,511 ~~~&  ~~8,365 \\
\tableline
total& 6,505 & 20,033 & 14,196~~ &~40,734 \\
\end{tabular}
\label{summary}
\end{table}

The London Stock Exchange consists of two parts: The completely
automated electronic downstairs market (SETS) and the upstairs market
(SEAQ).  The trading volume is split roughly equally between the two
markets.  We study the downstairs market because we have a record of
each action by each trader as it occurs.  In contrast, trades in the
upstairs market are arranged informally between agents, and are
printed later.  There are no designated market makers for SETS;
however, any member of the exchange is free to act as a market maker
by posting simultaneous bids and offers.  This should be contrasted
with the NYSE, for example, which has a designed specialist to trade
each stock\footnote{Another difference between the two markets is that
  clearing in the LSE is fully automated and instantaneous; in
  contrast, in the NYSE clearing is done manually, creating an
  uncertainty in response times.}  During the period we study the book
is fully transparent, i.e. all orders in the book are fully
revealed\footnote{In 2003 the LSE began to allow ``iceberg orders'',
  which contain a hidden component that is only revealed as the
  exposed part of the order is removed.}.

Trading begins each day with an opening auction.  There is a period
leading up to the opening auction in which orders are placed but no
transactions take place.  The market is then cleared and for the
remainder of the day (except for occasional exceptional periods) there
is a continuous auction.  We remove all data associated with the
opening auction, and analyze only orders placed during the continuous
auction.  

An analysis of the limit order placement shows that in our dataset
approximately $35\%$ of the effective limit orders are placed {\it
  inside the book} ($\pi > a(t)$ or $\pi < b(t)$).  $33\%$ are placed
at the best prices ($\pi = b(t)$ or $\pi = a(t))$, and $32\%$ are
placed inside the spread ($b(t) < \pi < a(t)$). This is roughly true
for all the stocks except for SHEL, for which the percentages are
$71\%$, $18\%$ and $11\%$, respectively\footnote{For Paris Stock
  Exchange from 1994 Biais, Hillion and Spatt \cite{Biais95} observed
  $42\%$ of the orders inside the spread, $23\%$ at the best, and
  $35\%$ inside the book.  We do not know why SHEL is anomalous,
  though it is worth noting that is the most heavily traded stock in
  the sample, a close proxy is also traded on the NYSE, and it has the
  second largest tail exponent among the $16$ stocks chosen for the
  study.}  Moreover for all the stocks the properties of buy and sell
limit orders are approximately the same.

In this dataset cancellation occurs roughly $32\%$ of the time at the
best price and $68\%$ of the time inside the book. This is quite
consistent across stocks and between the cancellation of buy and sell
limit orders. The
only significant deviation is once again SHEL, for which the percentages are
$14\%$ and $86\%$.

\section{Fluctuations in liquidity drive the tails}

The assumption that large price changes are caused by
large market orders is very natural.  A very large market order will
dig deeply into the limit order book, causing transactions at many
price levels, increasing the spread, and changing the midprice.
Surprisingly, this is {\it not} the cause of most large price changes.
Instead, as we will demonstrate in this section, most large price
changes are due to discrete fluctuations in liquidity, manifested by
gaps in filled price levels in the limit order book.  Large price
changes caused by large orders are very rare, and play an
insignificant role in determining the statistical properties of price
changes.

In this section we will focus on price changes caused by
market orders, and in Section~\ref{limitOrders} we will discuss price
changes due to limit orders and cancellations.

\subsection{Large returns are not caused by large orders}
\label{marketImpact}

We first demonstrate that most large returns are not caused by the
arrival of large market orders.  The probability density function of price
returns that are caused by market orders can trivially be written as
$p(r) = \int p(r | \omega) p(\omega) d\omega$, where $p(r)$,
$p(\omega)$ and $p(r | \omega)$ are the probability density functions
for returns $r$, market order size $\omega$, and returns given market
order size.  The conditional probability $p(r | \omega)$ characterizes the
response of prices to new orders, and can be viewed as the probability
density of market impacts, or alternatively, as characterizing the
distribution of liquidity for market orders.
When a market order of size $\omega$ arrives the midprice will move
if $\omega$ is larger than or equal to the volume at the matching best price
(i.e. the bid for sell market orders and the ask for buy market orders). 
In the limit of continuous prices we can trivially write 
\begin{equation}
p(r | \omega) = (1-g(\omega)) \delta(r) + g(\omega)f(r | \omega)
\end{equation}
where $\delta(.)$ is the Dirac delta function\footnote{The Dirac delta
  function $\delta(x)$ is defined so that $\int \delta(x) = 1$ over
  any domain that includes $0$ and $\int \delta(x) = 0$ otherwise.}
and $g(\omega)$ is the probability that the midprice moves as a
function of the order size $\omega$.  The function $f(r | \omega)$ is
the probability of a price shift $r$, conditioned on the midprice
moving in response to an order of volume $\omega$.  $g(\omega)$ and
$f(r | \omega)$ behave quite differently.  $g(\omega)$ depends
strongly on $\omega$.  For the LSE it scales roughly as $g(\omega) \sim
\omega^{0.3}$, about the same as the average market impact.  In
contrast, $f(r | \omega)$ is surprisingly independent of the volume
$\omega$, in the sense that the unconditional fluctuations in $r$
dominate the dependence on $\omega$.

\begin{figure}[ptb]
\begin{center} 
\includegraphics[scale=0.3,angle=-90]{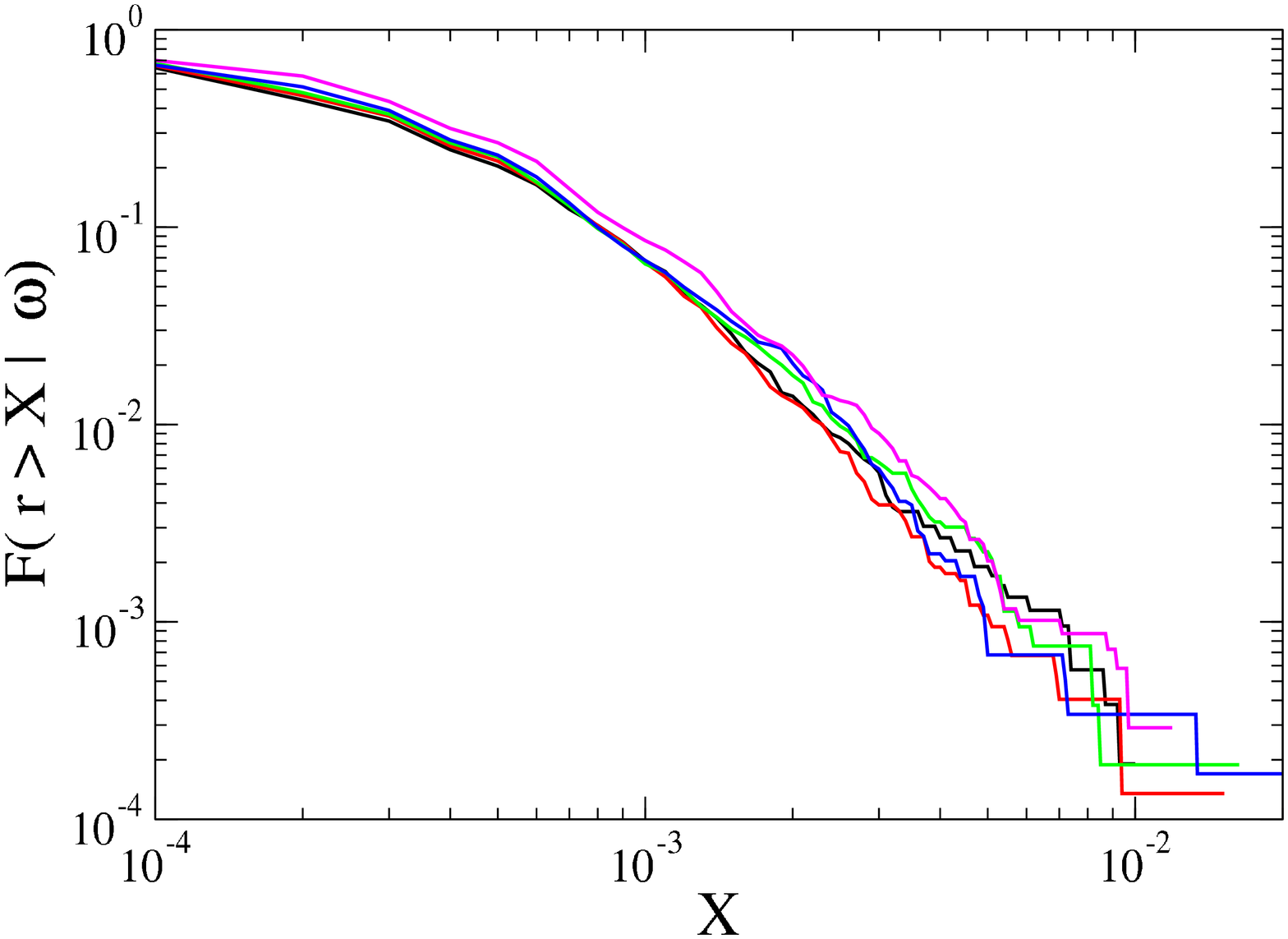}
\includegraphics[scale=0.3,angle=-90]{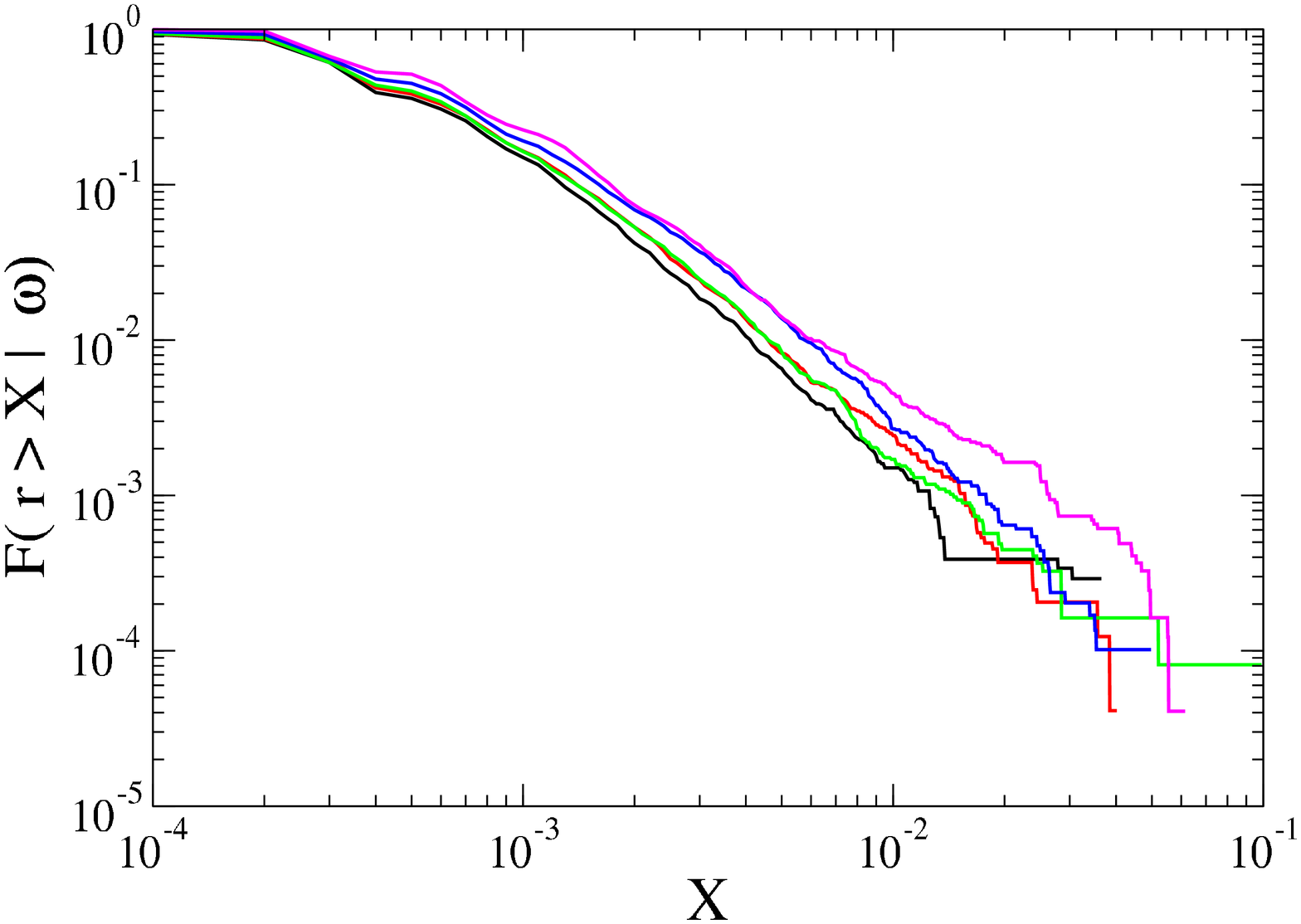}
 \caption{Dependence of returns on order size.  $F(r > X |\omega)$ is
   the probability of a return $r > X$ conditioned on the order size
   $\omega$ and on the fact that the price shift is nonzero.  Results
   shown are for buy orders, but similar results are seen for sell
   orders. The orders are sorted by size into five groups with roughly
   the same number of orders in each group.  Ranging from small orders
   to large orders, the curves are black, red, green, blue, and
   magenta.  In panel (a) we show the result for AZN and in panel (b)
   we show the average over the pool of $16$ stocks described in
   Table~\ref{summary}.  For the pooled data for each stock we
   normalize the order volume to the sample mean, and then combine the
   data. Each curve approximately approaches a power law for large $r$
   independent of $\omega$, illustrating that the key property
   determining large price returns is fluctuations in market impact,
   and that the role of the volume of the order initiating a price
   change is minor.}
 \label{pV} 
 \end{center}
\end{figure}

To demonstrate this in Figure~\ref{pV} we show the cumulative
probability for nonzero price returns conditioned on order size, i.e. $F(r>X |
\omega)=\int_X^{\infty} f(r | \omega) dr$ for several different ranges
of market order size $\omega$.
The orders are sorted by size into five groups with roughly the same
number of orders in each group.  The distributions for each range of
$\omega$ are roughly similar, both for individual stocks such as AZN,
and for the pool.  Each curve approximately approaches a power law for
large $r$ independent of $\omega$, illustrating that the key property
determining large price returns is fluctuations in market impact, and
that the role of the volume of the order initiating a price change is
minor.  For the pooled stock result, for large returns the curves for
large order size tend to be on top of those for small order size,
illustrating a weak dependence on order size, but this is relatively
small, and not visible in the results for individual stocks.  Although
we do not present them here, using the TAQ database we have obtained
similar results for a small sample of stocks traded in the NYSE,
suggesting that this behavior is not specific to the LSE\footnote{Note
  added in press: Based on five minute averages, Weber and Rosenow
  \cite{Weber04} also see that liquidity is the dominant effect, using
  data from NYSE and Island.}.  This is particularly
interesting given the significant differences in the structure of the
NYSE, and also because the TAQ data set includes upstairs trades.

\begin{figure}[ptb]
\begin{center} 
\includegraphics[scale=0.3,angle=-90]{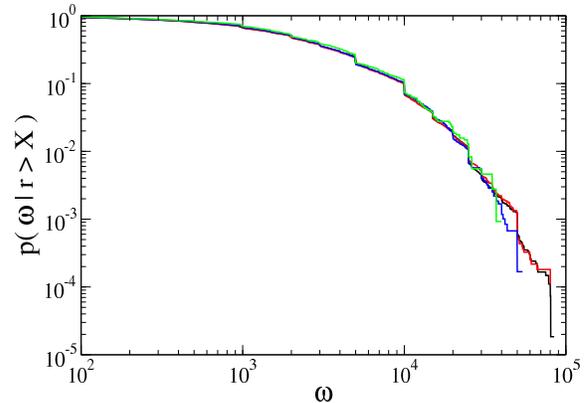}
 \caption{$p(\omega | r > X)$, the distribution of market order sizes
   conditioned on generating a return greater than $X$, for the stock
   AZN.  The values of $X$ correspond to the $50$ percentile (red),
   $90$ percentile (blue), and $99$ percentile (green) of the return
   distribution. The black line is the unconditional distribution
   $p(\omega)$.  The jumps are due to the tendency to place orders in
   round numbers of shares.}
 \label{condOrderSize}
\end{center}
\end{figure}

To reinforce this point, in Figure~\ref{condOrderSize} we show
$p(\omega | r > X)$, the probability of a market order of size
$\omega$ conditioned on the return being greater than a certain
threshold $X$.  We do this for the stock Astrazeneca (AZN) for several
different values of $X$, getting virtually the same curve independent
of $X$.
The distribution of order sizes that generate large returns are
essentially the same as those that generate small returns.  Similar
results are obtained for all the stocks in our sample.  Thus, it seems
that order size does not play an important role in generating large
returns.

Gabaix et al. \cite{Gabaix03} have recently proposed that large
returns can be explained by assuming the market impact density
function $p(r | \omega)$ is sharply peaked around a central
value\footnote{Analyses of several markets make it clear that the mean
  response of prices to orders varies considerably from market to
  market, and is not in general well characterized by a square root
  law \cite{Lillo03,Potters03,Farmer03,Farmer04}.  It is also worth
  noting that the volume distribution in Figure~\ref{condOrderSize}
  does not appear to have a power law tail. This is true of all the
  stocks in our sample.  In contrast, power law tails are observed for
  stocks in the NYSE \cite{Gopikrishnan00}, and this was also observed
  this to be true for upstairs data in the LSE (Lillo and Farmer,
  unpublished).  See also the discussion in \cite{Lillo03b}.}
$k\omega^{1/2}$, so that it can be approximated as a Dirac delta
function $\delta(r - k\omega^{1/2})$.  The result of Figure~\ref{pV}
makes it clear that the sharply peaked assumption is a poor
approximation -- the distribution is quite broad, in the sense that
the conditional distribution $p(r | \omega) \approx p(r)$, and $p(r)$
is not sharply peaked.

\subsection{Volume dependence of mean market impact}

From a variety of previous studies it is clear that the mean market impact
is an increasing function of $\omega$
\cite{Hasbrouck91,Hausman92,Farmer96,Torre97,Kempf99,Plerou02,Rosenow02,Evans02,Takayasu02,Lillo03,Potters03}.
How do we reconcile this with our claim that the distribution of
market impact is almost independent of volume?  The key is that the
main effect of changing order volume is to change the probability that
the price will move, with very little effect on how much it moves.
From Eq. (1) one easily obtains
\begin{equation}
E(r | \omega)= g(\omega) \int f(r |\omega)r~dr\propto  g(\omega) 
\end{equation}
where for the last proportionality relationship we have used the
result from Figure~\ref{pV} that $f(r |\omega)$ is almost independent
of $\omega$.  Thus, the expected price change scales like the
probability of a price change, a relationship that we have verified
for both the LSE and the NYSE.  However this variation is still small
in comparison with the intrinsic variation of returns; the mean market
impact of a very large order is less than the average size of the
spread, but the largest market impacts are often more than ten times
this large.

\subsection{Correlations between order size and liquidity}

One possible explanation of the independence of price response and
order size is that there is a strong correlation between order size
and liquidity.  There is an obvious strategic reason for this: Agents
who are trying to transact large amounts split their orders and
execute them a little at a time, watching the book, and taking
whatever liquidity is available as it enters.  Thus, when there is a
lot of volume in the book they submit large orders, and when there is
less volume, they submit small orders.  This effect tends to even out
the price response of large and small orders.  We will see that this
effect indeed exists, but it is only part of the story, and is not the
primary determinant of the behavior we observe here.

In fact, the unconditional correlation between market order size and
volume at the best is rather small.  For AZN, for example, it is about
$1\%$.  However, if we restrict the sample to orders that change the
midprice, the correlation soars to $86\%$.  The reason for this is
that for orders that do not change the price, there is essentially no
correlation between order size and volume at the best.  For the rarer
case of orders that do change the price, in contrast, most market
orders exactly match the volume at the best. As shown in
Table~\ref{bestvolume}, for the stocks in our sample, $86\%$ of the
buy orders that change the price exactly match the volume at the best
price.  (This is $85\%$ for sell orders.)
\begin{table}
\begin{tabular}{l|cc|cc}
&\multicolumn{2}{c|}{$\%$ of nonzero return }&\multicolumn{2}{c}{$\%$ of nonzero return}\\
&\multicolumn{2}{c|}{equal to first gap}&\multicolumn{2}{c}{with $\omega=V_{best}$}\\
tick &  sell& buy & sell& buy\\
\tableline
AZN  &  ~94.3   & 99.6&83.7 &    90.1 \\
BAA  &  ~95.9   & 99.1&86.2 &    87.0  \\
BLT  &  ~95.0    & 99.2& 85.8 &   85.6 \\
BOOT &  ~95.9 & 99.2&85.7 &    84.7  \\
BSY  &  ~94.3   & 99.7&85.0 &    88.0 \\
DGE  &  ~96.3  &  99.7&86.5  &  87.4 \\
GUS  &  ~95.8 &  99.5 &85.0 &    83.5  \\
HG.  &   ~95.9 &  99.5&85.8 &      83.0  \\
LLOY &  ~97.3 & 99.8& 88.4 &    88.6 \\
PRU  &  ~95.9 & 99.5 &85.5 &     78.1 \\
PSON &  ~93.1 & 99.6&81.2 &    86.2 \\
RIO  &  ~95.7 & 99.7 &84.6 &    86.4 \\
RTO  &  ~96.1 &  99.5&84.4 &     85.3  \\
RTR  &  ~93.0 & 99.7&83.5 &      85.4 \\
SBRY &  ~95.6 & 99.4 & 85.3&  83.2 \\
SHEL &  ~98.7 & 99.9&93.0 &   92.8 \\
\tableline
average& ~95.6 & 99.5 & 85.6&86.0 \\
\end{tabular}
\caption{Summary table of the percentage of the time that nonzero
  changes in the best prices are equal to the first gap (left) and
  that the market order volume $\omega$ exactly matches the volume at
  the corresponding best price $V_{best}$ (right). Assuming a
  Bernoulli process the sample errors are the order of $0.1-0.2\%$.}
\label{bestvolume}
\end{table}

\begin{figure}[ptb]
\begin{center} 
\includegraphics[scale=0.3,angle=-90]{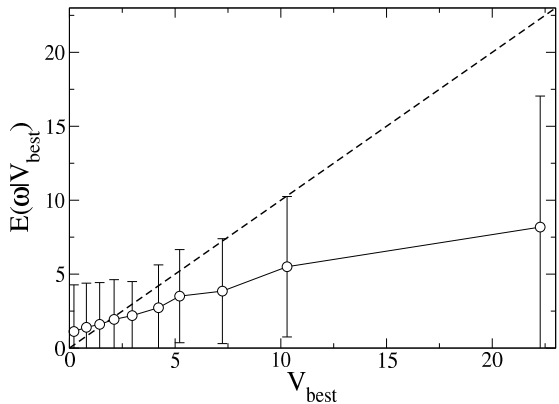}
 \caption{The dependence of order size on liquidity.  Liquidity is
   measured as the volume at the best price, which is binned into
   deciles with roughly equal numbers of events. The vertical axis
   shows the mean market order volume for each decile, and the
   horizontal shows the volume at the best. The units of both axes are
   thousands of shares.  The ranges shown around each point indicate
   one standard deviation (and are not standard errors). The dashed
   line has slope $1$ and serves as a point of comparison.}
 \label{orderSizeVsBest} 
 \end{center}
\end{figure}

The relationship between the volume of market orders and the best
price becomes more evident with a nonlinear analysis.
Figure~\ref{orderSizeVsBest} shows $E(\omega|V_{best})$, where
$\omega$ is the market order size and $V_{best}$ is the volume at the
corresponding best price.
We see in this figure that the expected order size is nonzero even for
the smallest values of $V_{best}$.  It grows monotonically with
$V_{best}$, but with a slope that is substantially less than one, and a
roughly concave shape.  This makes the nonlinear correlation between
order size and liquidity clear.  However, in the following section we
will see that the dependence of order size on liquidity is not strong
enough to substantially suppress large price fluctuations.

\subsection{Liquidity fluctuations drive price fluctuations}

In this section we demonstrate in concrete terms that price
fluctuations are driven almost entirely by liquidity fluctuations.  To
do this we study the virtual market impact, which is a useful tool for
probing the supply and demand curves defined by the limit order book.
Whereas the true market impact $p(r | \omega)$ tells us about the
distribution of impacts of actual market orders, as discussed in
Section~\ref{marketImpact}, the virtual market impact is the price
change that {\it would} occur at any given time if a market order of a
given size were to be submitted.  More formally, at any given time $t$
the limit orders stored in the order book define a revealed supply
function $S(\pi,t)$ and revealed demand function $D(\pi,t)$.  Let
$V(\pi,t)$ be the total volume of orders stored at price $\pi$.  The
revealed supply function is
\begin{equation}
S(\pi,t) = \sum_{i = a(t)}^{i=\pi} V(i,t)
\end{equation}
The revealed supply function is non-decreasing, and so for any fixed
$t$ has a well defined inverse $\pi(S,t)$.  The virtual market impact is
the price shift caused by a hypothetical order of size $S$, e.g. for
buy orders it is $\pi(S,t) - a(t)$.  The virtual market impact for sell
market orders can be defined in terms of the revealed demand in a
similar manner.  By sampling at different values of $t$, for any fixed
hypothetical order size we can create a sample distribution of virtual
market impacts.  This naturally depends on the sampling times, but
these can be chosen to match any given set of price returns.

\begin{figure}[ptb]
\begin{center} 
 \includegraphics[scale=0.3,angle=-90]{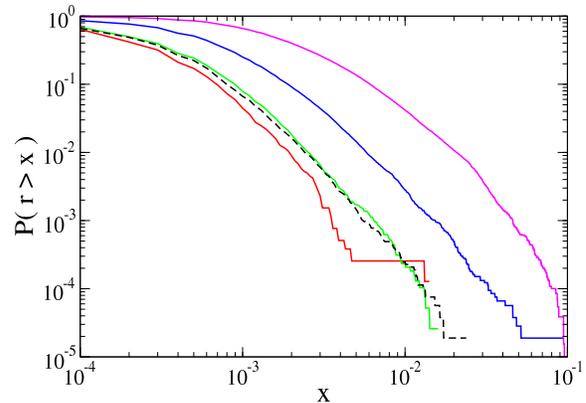}
 \caption{The cumulative probability distribution $P(r > x)$ of
   virtual market impacts for the stock AZN, based on samples of the
   limit order book taken just before the arrival of a market order.
   From left to right the continuous lines are the virtual market
   impact for hypothetical buy orders of size $\omega = 100$ shares
   (red), $\omega = 1,600$ shares (green), $\omega = 8,700$ shares
   (blue), and $\omega = 25,000$ shares (magenta); these correspond to
   the $0.1$, $0.5$, $0.9$, and $0.99$ quantiles of market order size.
   These are compared to the average distribution of returns (black
   dashed), which is very similar to the virtual market impact for the
   $0.5$ quantile.  Note that we have discarded cases in
   which the virtual market impact is undefined due to excessively
   large hypothetical market order volume, but these comprise only
   about $3\%$ of the events.}
\label{virtualImpact} 
\end{center}
\end{figure}

In Figure~\ref{virtualImpact} we show the cumulative distribution
of virtual market impacts for the stock AZN for several different values
of $D$, corresponding to different quantiles of market order size.
The cumulative distributions define a set of approximately parallel
curves.  They are shifted as one would expect from the fact that
larger hypothetical order sizes tend to have larger virtual market
impacts.  These curves are similar in shape to the distribution of
returns.  Most striking, for the median market order size, the curves
are almost identical.  We see similar results for all the stocks in
our sample.  This demonstrates quite explicitly that the distribution
of returns is determined by properties of the limit order book, and
that the typical price return corresponds to the price response to an
order of typical size.  The fact that the price distribution is so
close to the virtual market impact of a typical order shows that
correlations between order size and liquidity are not important in
determining price fluctuations.

\section{Granularity of supply and demand\label{granularity}}

At first sight, the behavior described in the previous section seems
baffling: How can market order size be so unimportant to price
response?  In this section we show how this is due to granularity of
revealed supply and demand, which causes large fluctuations in
liquidity.

\begin{figure}[ptb]
\begin{center} 
\includegraphics[height=1.2in, width=3.3in]{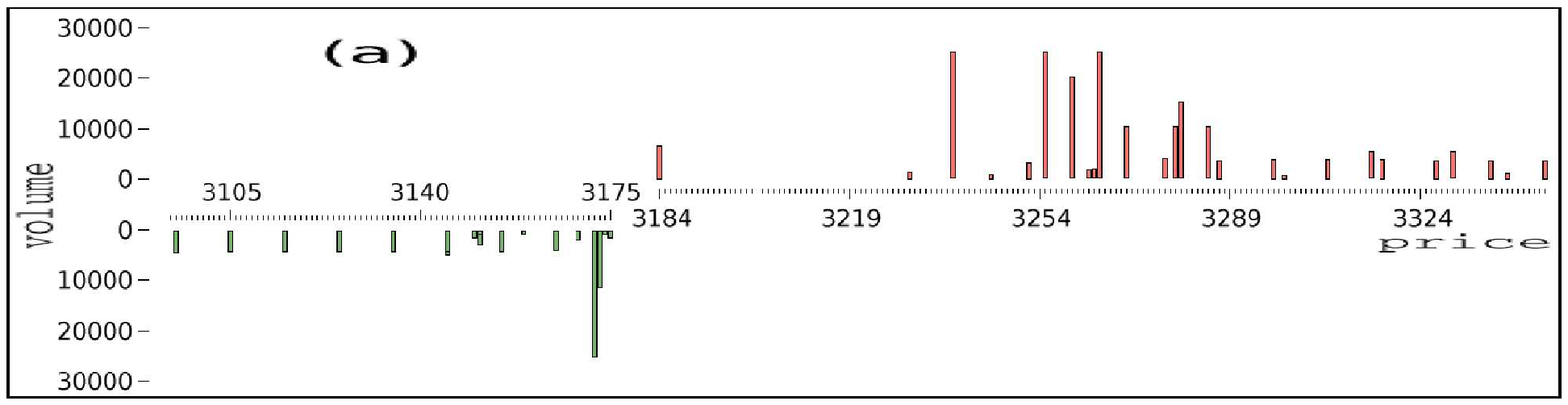}
\includegraphics[height=1.2in, width=3.3in]{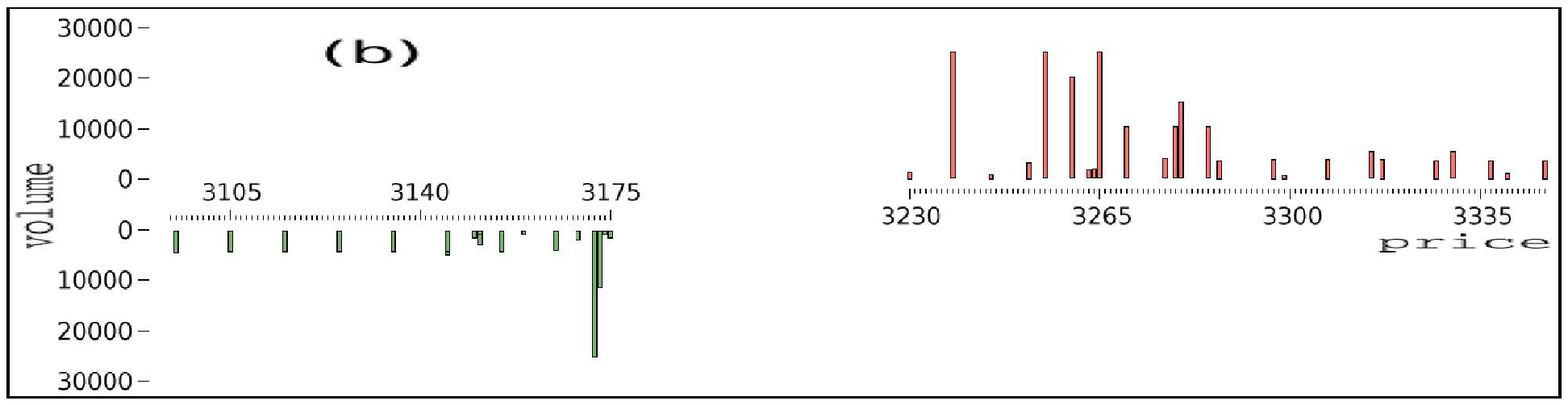}
 \caption{A typical configuration of the limit order book for AZN before and
   after a large price fluctuation.  The two panels plot the
   volume (in shares) of limit orders at each price level; sell limit
   orders are shown as positive, and buy limit orders as negative.  In
   panel (a) we see that there is a large gap between the best
   ask price and the next highest occupied price.  The arrival of a
   market order to buy removes all the volume at the best ask,
   giving the new limit order book configuration shown in panel (b),
   which has a much higher best ask price than previously.  
   }
\label{gapMovie} \end{center}
\end{figure}

The cause of this puzzling behavior is fluctuations in occupied price
levels in the limit order book.  In particular, one can define the
size of the {\it first gap} $g$ as the absolute difference between the
best log price $\pi_{best}$ and the log price of the next best quote,
$\pi_{next}$ as $g = |\log \pi_{best} - \log \pi_{next}|$.
In Figure~\ref{gapMovie} we show a typical set of events before and
after a large price fluctuation.
In panel (a) we see that there is a large first gap on the sell side
of the limit order book.  In panel (b) we see the configuration of the
book an event later, after a market order or exactly the same size has
removed all the volume at the best ask.  This results in a large
change in the midpoint price.  Thus, the large return simply reflects
a large gap inside the book that has been revealed by the removal of
the best ask.  We find that this is the typical behavior underlying
almost all large returns.

As already shown in Table~\ref{bestvolume}, typically about $85\%$ of
market orders that result in price changes exactly match the volume at
the best price.  Furthermore, as shown in the first two columns of
this table, when they do exceed the volume at the best price, it is
quite rare that they penetrate the next occupied level.  For buy
(sell) market orders, on average $95.5\%$ ($99.5\%$) of the nonzero
shifts in the best price are exactly equal to the first
gap. Interestingly there is a significant asymmetry between buy and
sell.  Traders act to minimize their transaction costs, so that jumps
of more than one gap are rare, particularly for sell orders.
(However, as we stressed earlier, the return distribution can be
generated by a constant order of median size -- this correlation is
interesting, but not essential).  Note that the trader initiating the
change {\it does not} pay a large spread -- that would only happen to
the next trader, if she were to immediately place a market
order\footnote{In general after a large shift in the bid or ask price,
  the next orders tend to be limit orders, but we have not yet been
  able to study the statistical properties of the sequence of
  subsequent events in detail.  See
  \cite{Cohen81,Biais95,Griffiths00,Hollifield02}, and
  \cite{Petersen94} for an empirical study of the role of quote size
  after a market order.}.

\begin{figure}[ptb]
\begin{center} 
\includegraphics[scale=0.3,angle=-90]{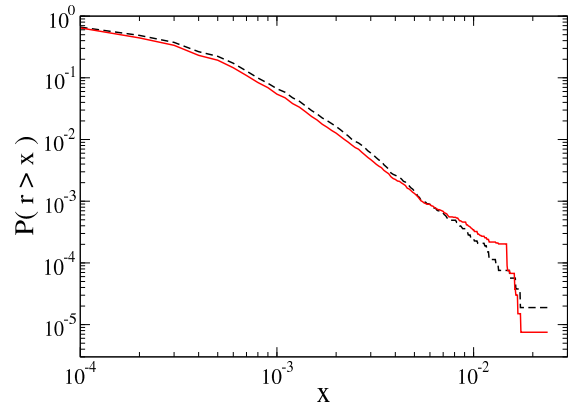}
\includegraphics[scale=0.3,angle=-90]{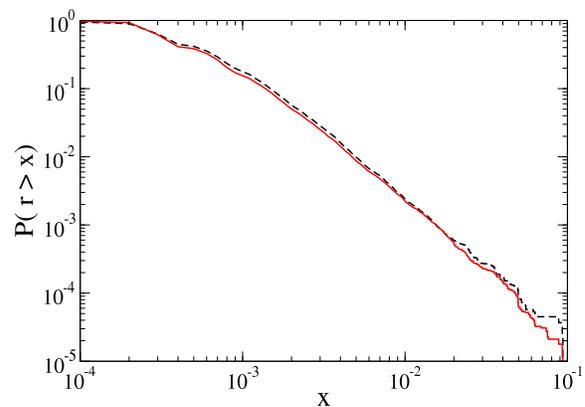}
 \caption{The cumulative distribution $P(g>x)$ of the size of first
   gaps $g$ (red continuous line), compared to the cumulative
   distribution of returns generated by market orders $P(r>x)$ (black
   dashed line). Panel (a) refers to buy market orders for AZN,
   in double logarithmic scale to highlight the tail behavior.  The
   two distributions are very similar. The result is even more
   impressive when we consider the average over the $16$ stocks
   described in Table~\ref{summary} (Panel (b))}
\label{bestHole} \end{center}
\end{figure}

This is by far the most common scenario that generates large
price changes.  In Figure~\ref{bestHole}(a) we compare the distribution
of the first gaps to the distribution of price returns for the stock AZN.
We see that the distributions are very similar.  For the first gap
size the tail index $\alpha =2.52\pm 0.07$, and for the return
distribution $\alpha = 2.57\pm0.08$, showing that the scaling
behaviors are similar.  However, the similarity is not just evident in
the scaling behavior -- the match is good throughout the entire range,
illustrating that most of the large price changes are caused by events
of this type.  Panel (b) of Figure~\ref{bestHole} shows the same
comparison for the pool of $16$ stocks.  The agreement in this case is
even more striking\footnote{We have also studied the distribution of
  higher order gaps.  Moving away from the best price $\pi_0$, the
  $n^{th}$ order gap for $n = 1, 2, \ldots$ can be recursively defined
  as $g_n = | \log \pi_{n-1} - \log \pi_{n} |$, where $\pi_n$ is the $n^{th}$
  occupied price level.  Interestingly, we find that the tail behavior
  of higher order gaps is the same as that of $g_1$.}.

\begin{figure}[ptb]
\begin{center} 
\includegraphics[scale=0.3,angle=-90]{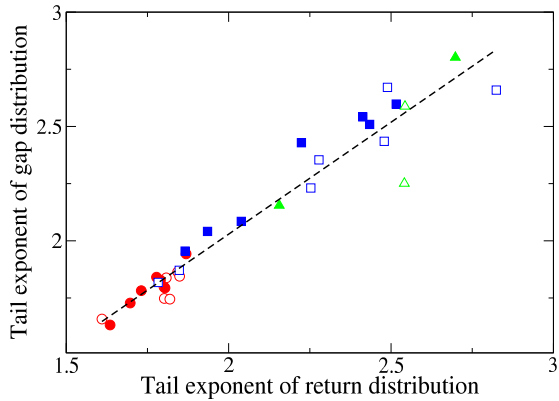} 
 \caption{A comparison of tail exponents $\alpha$ for returns
   (horizontal axis) vs. tail exponents for the first gap (vertical
   axis).  The first gaps are sampled immediately preceding the market
   order that triggers the return.  This is done for all the stocks in
   Table~\ref{summary}, and includes both the buy and sell side of the
   limit order book.  We use red circles for low liquidity stocks,
   blue squares for medium liquidity stocks, and green triangles for
   high liquidity stocks.  Empty symbols refer to sell market orders
   and filled symbols to buy market orders.  The black dashed line is
   the linear regression.  We see that there is a clear positive
   relationship between the number of orders and the tail exponent.}
\label{scatterRtn} \end{center}
\end{figure}

To demonstrate that the correspondence in the above figure is not just
a coincidence for AZN, we have computed the tail exponents for returns
and first gap size for all $16$ stocks in our data set.  We do this
using a Hill estimator \cite{Hill75} by considering the largest
$\sqrt{n}$ returns, where $n$ is the size of the sample.  The results
are shown in Figure~\ref{scatterRtn}, where we plot the tail exponents
of returns against those of first gaps.
The points cluster tightly along the diagonal, making it clear that
there is a strong positive correlation ($R^2=0.93$) between the gap
tail exponents and the return tail exponents; a least squares linear
fit has a slope of $0.98 \pm 0.05$, and we are unable to reject the
null hypothesis that the tail exponents of the gaps and returns are
drawn from the same distribution.

\begin{figure}[ptb]
\begin{center}
\includegraphics[scale=0.3,angle=-90]{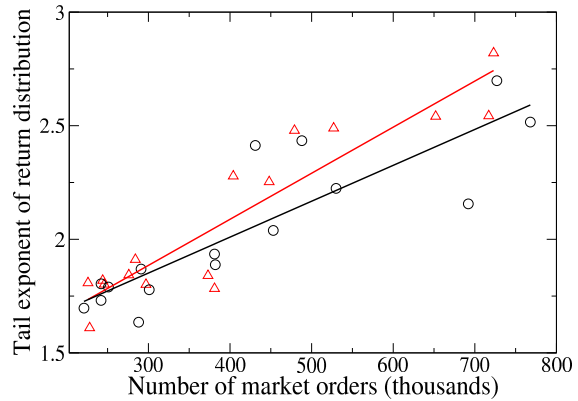}
 \caption{Dependence of the tail exponent of price changes on the
   number of market orders, for the data of Figure~\ref{scatterRtn}.
   Black circles are for positive returns (caused by buy market
   orders), and red triangles are for negative returns (caused by sell
   market orders).  The steeper red curve is the fit to negative
   returns and the black curve the fit to positive returns.}
  \label{tailVsEvents} \end{center}
\end{figure}

It is worth noting that lightly traded stocks tend to have a smaller
tail exponent than heavily traded stocks.  This is already clear in
Figure~\ref{scatterRtn}, where we have used a color coding to identify
stocks of different volume.  To investigate this more quantitatively,
in Figure~\ref{tailVsEvents} we plot the tail exponent of price
changes against the number of market orders for each stock in the
sample.  While the relationship is noisy, there is a clear positive
trend; the slopes of the linear fits to positive and negative returns
are both highly statistically significant. The tail exponents vary
from about $1.6$ to $2.8$, whereas the error bars are more than a
factor of ten smaller than the range of
variation\footnote{Table~\ref{hill} gives error bars for tail exponent
  estimations based on absolute returns (including both buy and sell
  events), while Figures~\ref{bestHole} and \ref{scatterRtn} are based
  on buy and sell returns taken separately, but the error bars are
  comparable.}.  This suggests that the tail exponent is not
universal.  Not surprisingly, stocks that are more heavily traded tend
to display less extreme risk than stocks that are lightly traded.

\section{Limit orders and cancellations}\label{limitOrders}

We have so far considered only returns caused by market orders.  In
this section we will discuss returns caused by limit orders and
cancellations, and show that they are statistically indistinguishable
from those caused by market orders.

\begin{figure}[ptb]
\begin{center} 
\includegraphics[scale=0.3,angle=-90]{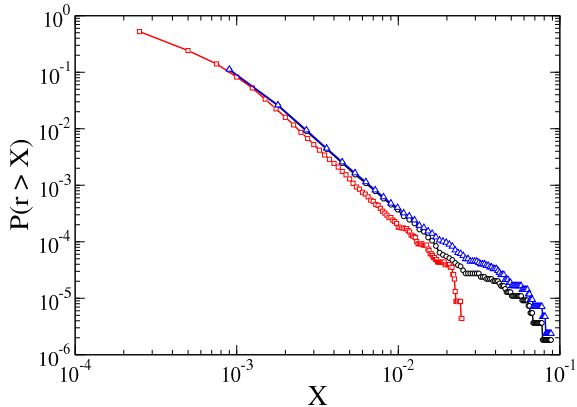}
 \caption{A comparison of the distribution of absolute returns
   (including both buy and sell orders) caused by limit orders, market
   orders, and cancellations for AZN.  The black circles are for
   market orders, the blue triangles are for limit orders, and the red
   squares are for cancellations.}
\label{loReturns} 
\end{center}
\end{figure}

The distribution of returns caused by limit orders, market orders, and
cancellations for AZN is shown in Figure~\ref{loReturns}.
The distributions are quite similar.  To make the analysis more
quantitative, we compute the tail exponents for the $16$ stocks in the
sample.  Table~\ref{hill} shows the Hill estimates for returns caused
by market orders, limit orders and cancellations.  The table shows
clearly that the tail exponents for the three distributions are very
close: For all $16$ stocks the estimated tail exponents are well
within the $95\%$ confidence intervals.  On the other hand
Table~\ref{hill} shows that there is a much larger variation across
stocks.

\begin{table}
\caption{Hill estimator of the return caused by market orders, limit
  orders and cancellations. We analyzed the top percentile and give
  $95\%$ confidence intervals.  The confidence intervals
  assume the samples are uncorrelated, and so may be over-optimistic.}
\begin{tabular}{l|ccc}
tick &  Market orders& Limit orders & Cancellations\\
\tableline
AZN& $2.57 \pm0.08 $&$2.56 \pm0.07 $&$2.57 \pm0.06$\\
BAA& $1.77 \pm0.06 $&$1.77 \pm0.08 $&$1.76 \pm0.08$ \\
BLT &$1.8 \pm0.1 $&$1.8 \pm0.1 $&$1.8 \pm0.1$\\
BOOT &$1.80 \pm0.07 $&$1.80 \pm0.07 $&$1.79 \pm0.08$\\
BSY &$2.28 \pm0.04 $&$2.25 \pm0.06 $&$2.26 \pm0.06$\\
DGE &$2.31 \pm0.06 $&$2.31 \pm0.07 $&$2.31 \pm0.07$\\
GUS &$1.80 \pm0.09 $&$1.80 \pm0.07 $&$1.80 \pm0.07$\\
HG. &$1.7 \pm0.1 $&$1.7 \pm0.1 $&$1.66 \pm0.07$\\
LLOY& $2.72 \pm0.03 $&$2.72 \pm0.04 $&$2.72 \pm0.05$\\
PRU &$2.20 \pm0.06 $&$2.20 \pm0.07 $&$2.20 \pm0.06$\\
PSON &$1.9 \pm0.1 $&$1.9 \pm0.1 $&$1.91 \pm0.09$\\
RIO &$1.83 \pm0.09 $&$1.82 \pm0.08 $&$1.83 \pm0.07$\\
RTO &$1.73 \pm0.09 $&$1.73 \pm0.08 $&$1.73 \pm0.07$\\
RTR &$2.45 \pm0.04 $&$2.44 \pm0.05 $&$2.45 \pm0.03$\\
SBRY &$1.89 \pm0.07 $&$1.88 \pm0.07 $&$1.88 \pm0.07$\\
SHEL &$2.62 \pm0.04 $&$2.61 \pm0.07 $&$2.62 \pm0.04$\\
\end{tabular}
\label{hill}
\end{table}

Why are the distributions of different events so similar?  The
correspondence between returns caused by market orders and returns
caused by cancellations is not surprising. In terms of the effect on
best prices, removal of the volume at the best price by cancellation
is equivalent to removal by a market order, in both cases creating a
price change equal in size to the first gap.  However, for a limit
order this is not so obvious: A limit order that falls inside the
spread decreases the spread, and causes a price change in the opposite
direction from market orders and cancellations.  To investigate this
we have studied the distribution of the spread, which also appears to
be a power law, but with a larger tail exponent.  At this point the
reason for the close correspondence between the returns generated by
limit and market orders remains unexplained.

\section{Price changes on longer timescales\label{aggregation}}

One can naturally ask whether the microscopic event level analysis we
have presented here explains the statistical properties of price
changes over longer and/or fixed time horizons.  The timescale for the
analyses we have presented here is quite fast: For Astrazeneca, for
example, within the period of this dataset there are on average about
eight market orders every five minutes, but due to highly uneven rates
of event arrival, it is not uncommon that many events arrive within
the same second.  It is natural to question whether events on this
timescale reflect the properties of longer timescales, e.g. on the
daily timescale of many other studies.  Given that order arrival is
highly clustered in time, it is also natural to ask whether a
description in event time also provides an explanation in real time.
We now address both of these questions.

\begin{figure}[ptb]
\begin{center} 
\includegraphics[scale=0.3,angle=-90]{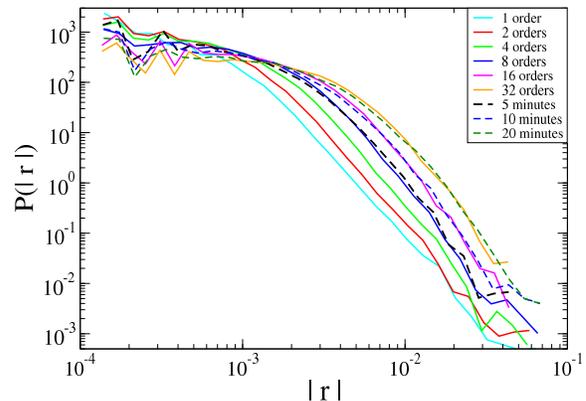}
 \caption{Price returns aggregated on different timescales and for
   different number of trades.  The price return is defined as $\log
   m(t) - \log m(t - \tau)$, where $\tau$ is from left to right $1$,
   $2$, $4$, $8$, $16$ and $32$ market orders (continuous lines) and
   $5$, $10$, and $20$ minutes (dashed lines).   The data is for AZN, based on absolute returns (both buy and
   sell).}
\label{multipleEventRtn} \end{center}
\end{figure}

The permanence of these price movements can be seen through the
continuity between price returns on the single event scale and the
multiple event scale.  Figure~\ref{multipleEventRtn} shows the return
density function for AZN for $1$, $2$, $4$, $8$, $16$, and $32$ market
order arrivals, as well as for $5$, $10$, and $20$ minute timescales.
Each curve consists of a non-power law central behavior crossing over
to an approximate power law tail behavior.  We see that the tails of
all the distributions are quite similar, though the crossover point
from the central behavior to the tail behavior increases for longer
timescales.  The regularity of the movement in the crossover point as
we vary the timescale, and the similarity of the tail behavior across
a range of timescales, indicates that the tail properties on the
single event scale are reflected on larger scales (at least as large
as $20$ minutes).  Also, this figure makes it clear that the behavior
in real time is essentially the same as the behavior in event time.
We see similar behavior for all the stocks in our sample.  Since other
studies have shown the continuity between timescales from $10$ minutes
to as long as a month \cite{Longin96,Plerou99,Mantegna99}, it seems
clear that an explanation at the level of single events is
sufficient\footnote{We have studied the response to individual events,
  and see some evidence for mean reversion of large price changes.
  However, the results of Figure~\ref{multipleEventRtn} make it clear
  that these are not sufficient to undo the initial change.}.  This is
particularly striking, given that on a longer timescale other
processes may become important, such as order splitting.
Figure~\ref{multipleEventRtn} suggests that for understanding the
statistical properties of prices these can be neglected.

The fact that we observe essentially identical distributions when we
aggregate over individual transactions or over fixed time rules out
any explanation of fat tails for price returns based on a subordinated
random process, as was suggested by Clark~\cite{Clark73}.  While
fluctuations in the number of events in a given length of time might
be quite important for other phenomena, such as clustered volatility,
they are clearly not important in determining the price return
distribution.

\section{Correlations in occupied sites}

A natural hypothesis about the origin of the large gap distributions
involves fluctuations in the number of occupied sites in the
order book.  If there are only a few orders in the book, so that it is
mostly empty, we naturally expect large gaps to exist.  Perhaps the
approximate power for the gap distribution can be explained by a
similar power law in the number of occupied sites?  In this section we
show that this is not the case, but rather that the power law behavior
depends on correlations in occupied price levels.

\begin{figure}[ptb]
\begin{center} 
\includegraphics[scale=0.3,angle=-90]{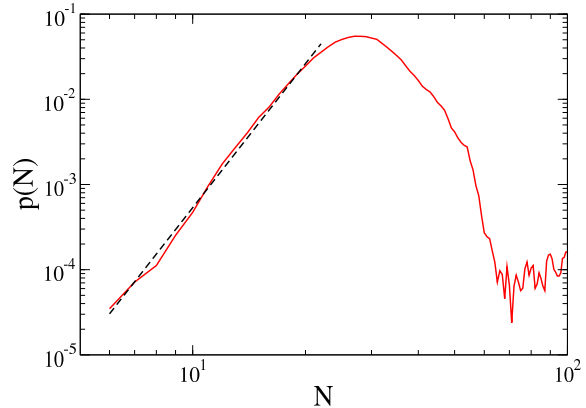}
\includegraphics[scale=0.3,angle=-90]{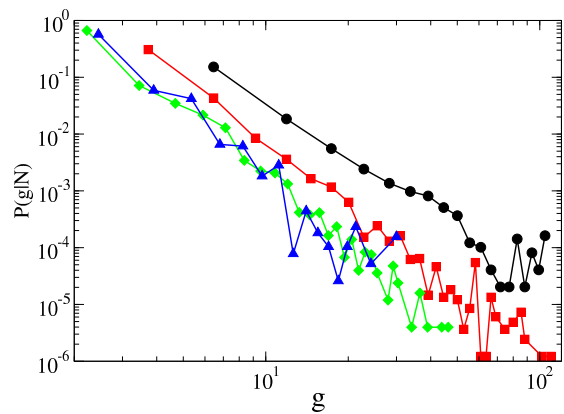}
 \caption{The effect of the finite number of orders for the stock AZN.
   Panel (a) shows the unconditional distribution of the number of
   occupied price levels for buy orders, plotted on double logarithmic
   scale. The dashed line is the best fit of the low $N$ region with a
   functional form $p(N)\propto N^{\beta}$.  Panel (b) shows the
   probability density of first gap sizes conditional on the number of
   occupied price levels, $p(g | N)$.  We divide the sample in four
   subsamples according to the value of $N$. Specifically, we have
   $0<N\le15$ (black circles), $15<N\le 30$ (red squares), $30<N\le
   45$ (green diamonds), and $45<N\le 60$ (blue
   triangles). Surprisingly, the first gap size shows approximate
   power law behavior even when the number of occupied sites is very
   large.}
\label{gapDist} \end{center}
\end{figure}

Interestingly, we do find evidence of power law scaling in the number
of occupied price levels. Figure~\ref{gapDist} (a) shows the
probability density function $p(N)$, where $N$ is the number of
occupied price levels in the order book at any given time. The low $N$
region is well fit by a power-law $p(N)\sim N^\beta$ over about two
decades.

However, the fluctuations in the number of orders stored in the book
cannot explain the power law behavior of prices, because the tail
exponent of the power law behavior of the gap distribution near zero
is much too large.  If we assume that the location of order deposition
is uncorrelated, then for order arrival on a bounded domain gap size
should be proportional to $1/N$.  But with $p(N) \sim N^\beta$ and $g
\sim 1/N$, $p(g) \sim g^{-(\beta + 2)}$.  For $\beta \approx 5.5$ this
results in $p(g) \sim g^{-7.5}$.  In contrast, the empirically
observed scaling exponent for the density for AZN is about $3.5$.

Surprisingly, the distribution of gap size shows approximate power law
scaling, independent of $N$.  This is illustrated for AZN in
Figure~\ref{gapDist}(b), where we plot the probability of gap size
conditioned on $N$, $p(g | N)$, for several different values of $N$.
The same approximate power law scaling behavior is seen independent of
$N$.  The approximate power law behavior is evident even with as many
as $60$ occupied sites on one side of the book.  Considering that it
is rare for orders to be placed more than $100$ price ticks away from
the best price, this illustrates that occupied sites display
nontrivial correlations, which are essential for explaining large
price fluctuations.  This conclusion is reinforced by studies we have
done comparing the real distribution of occupied levels with models
based on IID order placement, which do not reproduce the power law
behavior.

\section{Conclusions}

For the London Stock Exchange, we have shown that large fluctuations
in prices are unrelated to large transactions, or to the placement of
large orders.  Instead, large price fluctuations occur when there are
gaps in the occupied price levels in the limit order book.  Large
changes occur when a market order removes all the volume at the best
price, creating a change in the best price equal to the size of
the gap.  

At a higher level, these results demonstrate that large price changes
are driven by fluctuations in liquidity.  There are times when the
market absorbs changes in supply and demand smoothly, and other times
when a small change in supply or demand can result in a very large
change in the price.  This is due to the fact that supply and demand
functions are not smooth, but rather have large, irregular steps and
jumps.  The market is granular, due to the presence of only a finite
number of occupied price levels in the book. This is what in physics
is called a finite size effect.  Even for an active stock such as AZN,
the number of occupied price levels on one side of the book at any
given time is typically about $30$, and so the system is far from the
continuum limit.  However, we have shown that this is not a simple
matter of fluctuations in the number of occupied price levels; while
there is an approximate power law in the limit $N \rightarrow 0$ in
the frequency for occupying $N$ price levels, this is not sufficient
to explain fluctuations in prices.  Instead, the power law in gaps
persists even when the number of occupied levels is quite high,
reflecting nontrivial correlations in the positions of orders sitting
in the book.

The empirical results that we have presented here raise as many
questions as they answer.  In particular, what is responsible for the
approximate power law distribution of gap sizes, and why do limit
orders have the same tail exponents as market orders and
cancellations?  We have done some modeling to address this question.
In particular, we have modified the statistical model of order flow
introduced by Daniels et al. \cite{Daniels03,Smith03} to include
nonuniform order placement, more closely reflecting the empirical
order placement distribution \cite{Bouchaud02,Zovko02}, which has an
approximate power law tail.  For some parameter values our simulations
reproduce the power law tails of the gap distribution (and hence of
returns), but since we do not yet understand the necessary and
sufficient conditions for this, we do not include these results here.

The work presented here suggests that it is important to properly
model market institutions.  However, at this point it is not clear
how these results would change for a different market institution.
Fat tails are also observed in markets, such as the London Metals
Exchange, that follow very different exchange mechanisms\footnote{One
  of us (JDF) did an unpublished study in 1995 of daily data for
  industrial metals such as copper, zinc, iron, nickel, etc. traded on
  the London Metals Exchange, whose exchange mechanism resembles a
  Walrasian market.  All of these have large kurtosis, indicating fat
  tails.}.  While the LSE follows a continuous double auction, it is
our belief that key elements are likely to persist with other market
mechanisms.  In particular, we hypothesize that large price
fluctuations in any market are driven by liquidity fluctuations, and
that the granularity of fluctuations in supply and demand remains the
key factor underlying extreme price fluctuations.

These results are important because they reveal the detailed mechanism
through which prices display (at least approximate) power law
fluctuations.  They suggest that many previous models that claimed to
explain this phenomenon were misdirected, and provide strong
constraints on future models.  The tail exponent of large price
changes appears to depend on parameters of the market: More lightly
traded markets (with lower event rates) tend to display fatter tails,
with more extreme risks.  This has important practical importance
because it gives some understanding of what determines financial
risks, and gives some clues about how to reduce them.

\acknowledgments We would like to thank the James S. McDonnell
Foundation for their Studying Complex Systems Research Award, Credit
Suisse First Boston, McKinsey Corporation, Bob Maxfield, and Bill
Miller for supporting this research. We would also like to thank
Marcus Daniels for technical support.

\bibliographystyle{plainnat}
\bibliography{../bib} 
 
\end{document}